\documentclass[superscriptaddress, floatfix, reprint]{revtex4-2}
\setcitestyle{super,open={},close={}}

\usepackage{hyperref, graphicx, xspace}
\hypersetup{hidelinks, allcolors=blue, colorlinks=true}

\usepackage{multirow,booktabs,siunitx,natmove,threeparttable}
\AtBeginDocument{\RenewCommandCopy\qty\SI}
\DeclareSIUnit{\angstrom}{\textup{\AA}}
\DeclareSIUnit{\hartrees}{hartrees}
\DeclareSIUnit{\bohr}{bohr}

\usepackage[version=4]{mhchem}
\usepackage{cleveref}

\usepackage[acronym]{glossaries-extra}
\setabbreviationstyle[acronym]{long-postshort-user}
\glsdisablehyper

\newacronym{sa-casscf}{SA-CASSCF}{state-averaged complete active space self-consistent field}
\newacronym{nevpt2}{NEVPT2}{$n$-electron valence state second-order perturbation theory}
\newacronym{qdnevpt2}{QD-NEVPT2}{quasidegenerate $n$-electron valence state second-order perturbation theory}
\newacronym{mcpdft}{MC-PDFT}{multiconfiguration pair-density functional theory}
\newacronym{caspt2}{CASPT2}{complete active space second-order perturbation theory}
\newacronym{ms-pdft}{MS-PDFT}{multi-state pair-density functional theory}
\newacronym{ms-caspt2}{MS-CASPT2}{multi-state complete active space second-order perturbation theory}
\newacronym{xms-caspt2}{XMS-CASPT2}{extended multi-state complete active space second-order perturbation theory}
\newacronym{lpdft}{L-PDFT}{linearized pair-density functional theory}
\newacronym{ktsh}{$\kappa$TSH}{curvature-driven trajectory surface hopping}
\newacronym{kcsdm}{$\kappa$CSDM}{curvature-driven coherent switching with decay of mixing}
\newacronym{cc2}{CC2}{second-order coupled cluster}
\newacronym{cc3}{CC3}{third-order coupled cluster}
\newacronym{adc2}{ADC(2)}{second-order algebraic diagrammatic construction}
\newacronym{tddft}{TD-DFT}{time-dependent density functional theory}
\newacronym{cis}{CIS}{configuration interaction singles}
\newacronym{ri-adc2}{RI-ADC(2)}{resolution-of-the-identity second-order algebraic diagrammatic construction}
\newacronym{mrci}{MRCI}{multireference configuration interaction}
\newacronym{mrcisdq}{MRCISD+Q}{multireference configuration interaction singles and doubles with Davidson size-extensivity correction}
\newacronym{ipea}{IPEA}{ionization-potential--electron-affinity}
\newacronym{abld}{aBLD}{absolute bond-length difference}

\usepackage{physics, calrsfs, mathtools, bm}
\DeclareMathAlphabet{\pazocal}{OMS}{zplm}{m}{n}

\newcommand*{\PySCF}{\textsc{PySCF}\xspace}
\newcommand*{\PySCFforge}{\textsc{PySCF-forge}\xspace}
\newcommand*{\OpenMolcas}{\textsc{OpenMolcas}\xspace}
\newcommand*{\mrh}{\textsc{mrh}\xspace}
\newcommand*{\geomeTRIC}{\textsc{geomeTRIC}\xspace}
\newcommand*{\libxc}{\textsc{libxc}\xspace}
\newcommand*{\libcint}{\textsc{libcint}\xspace}
\newcommand*{\SHARC}{\textsc{SHARC}\xspace}

\begin{document}

\title{Semiclassical Nonadiabatic Molecular Dynamics Using Linearized Pair-Density Functional Theory}

\author{Matthew R. Hennefarth}
\affiliation{Department of Chemistry and Chicago Center for Theoretical Chemistry, University of Chicago, Chicago, IL 60637, USA}

\author{Donald G. Truhlar} \email[corresponding author: ]{truhlar@umn.edu}
\affiliation{Department of Chemistry, Chemical Theory Center, and Minnesota Supercomputing Institute, University of Minnesota, Minneapolis, MN 55455-0431, USA}

\author{Laura Gagliardi} \email[corresponding author: ]{lgagliardi@uchicago.edu} 
\affiliation{Department of Chemistry, Pritzker School of Molecular Engineering, The James Franck Institute, and Chicago Center for Theoretical Chemistry, University of Chicago, Chicago, IL 60637, USA}
\affiliation{Argonne National Laboratory, 9700 S. Cass Avenue, Lemont, IL 60439, USA}

% \date{\today}
\date{September 13, 2024}

\begin{abstract}
	Nonadiabatic molecular dynamics is an effective method for modeling nonradiative decay in electronically excited molecules. Its accuracy depends strongly on the quality of the potential energy surfaces, and its affordability for long direct-dynamic simulations with adequate ensemble averaging depends strongly on the cost of the required electronic structure calculations. Linearized pair-density functional theory (L-PDFT) is a recently developed post-self-consistent field multireference method that can model potential energy surfaces with an accuracy similar to expensive multireference perturbation theories but at a computational cost similar to the underlying multiconfiguration self-consistent field method. Here we integrate the \SHARC dynamics and \PySCF electronic structure code to utilize L-PDFT for electronically nonadiabatic calculations and use the combined programs to study the photoisomerization reaction of \textit{cis}-azomethane. We show that L-PDFT is able to successfully simulate the photoisomerization and yields results similar to the more expensive extended multi-state complete active space second-order perturbation theory. This shows that L-PDFT can model internal conversion, and it demonstrates its promise for broader photodynamics applications.
\end{abstract}

\maketitle

\newpage

Nonadiabatic molecular dynamics is a valuable tool for modeling nonradiative decay pathways in molecules. The quality of the dynamics depends on the accuracy of the potential energy surfaces on which the nuclei move, but obtaining good accuracy for excited-state potential energy surfaces is difficult due to the inherently multiconfigurational nature of excited-state wave functions and the need for consistent treatment of close-lying states. \Gls{sa-casscf} \cite{RoosComplete1980, RuedenbergMCSCF1979, RoosComplete1987} is a widely used multireference method that generates qualitatively accurate and consistent wave functions for a set of orthogonal states, but it does not yield quantitatively accurate potential energy surfaces. Multi-state multireference perturbation methods such as \gls{ms-caspt2} \cite{AnderssonSecond1990,FinleyMulti1998} or \gls{qdnevpt2} \cite{AngeliIntroduction2001, AngeliQuasidegenerate2004} use SA-CASSCF wave functions as a starting point to achieve better quantitative accuracy; however, they are computationally expensive. This limits their use to molecular dynamics of small molecular systems with small active spaces, short-timescale dynamics, or limited ensemble averaging.

\Gls{mcpdft} \cite{LiManniMulticonfiguration2014, GhoshCombining2018, ZhouElectronic2022} is an alternative post-SCF method that can yield accurate energies similar to \gls{caspt2} \cite{HoyerMulticonfiguration2016} or \gls{nevpt2} \cite{KingLarge2022} but at a fraction of the computational cost. Additionally, \gls{mcpdft} has been shown to agree with \gls{ms-caspt2} \cite{FinleyMulti1998} for intersystem crossing dynamics (population transfer to states of different spin symmetry) of thioformaldehyde, whereas \gls{sa-casscf} overestimated the population transfer \cite{MaiInfluence2019, ZhangDirect2021, CalioNonadiabatic2022}. \Gls{mcpdft} computes the total energy of a multiconfigurational wave function using a functional of the electron density and on-top pair density; however, it is a single-state method (it calculates the energy of each state independently of the other states). Single-state methods cannot accurately model the potential energy surfaces near regions of strong nuclear-electronic coupling such as at conical intersections, locally avoided crossings, or when many electronic states lie close in energy (as is common in actinides). As such, it is not an appropriate method for modeling internal conversion processes (population transfer to states of the same spin symmetry).

\Gls{lpdft} \cite{HennefarthLinearized2023} is a recently developed multi-state extension of \gls{mcpdft} that can properly model potential energy surfaces near regions of strong nuclear-electronic coupling. \Gls{lpdft} defines an effective Hamiltonian operator that is a functional of the model-space averaged density and on-top pair density with the property that its eigenvalues are linear approximations to the \gls{mcpdft} energies of the states. It is an example of a multi-state method in which diagonalization of the effective Hamiltonian operator within the model space yields potential energy surfaces with the correct topology near conical intersections and locally avoided crossings. \Gls{lpdft} has recently been shown to be as accurate as \gls{mcpdft} and \gls{nevpt2} at predicting vertical excitation energies for over 400 excitations in the QUESTDB database \cite{HennefarthLinearized2023a}, it is slightly faster than \gls{mcpdft}, it does not suffer from the intruder-state problem \cite{IijimaNote1985}, it does not require the iterative solution of a large set of linear perturbation-theory equations \cite{FinleyMulti1998}, and it does not require iterative steps to find an intermediate basis to construct an effective model-space Hamiltonian \cite{BaoCompressedState2020}. All of these attributes make \gls{lpdft} a very promising method for modeling excited-state dynamics. Recent development of \gls{lpdft} analytic nuclear gradients \cite{HennefarthAnalytic2024} as well as curvature-driven approximations to the time-derivative coupling for dynamics calculations \cite{BaeckPractical2017, ShuNonadiabatic2022, DoCasalFewest2022, ShuGeneralized2024} allows one to use \gls{lpdft} for modeling internal conversion processes using semiclassical nonadiabatic molecular dynamics methods such as \gls{ktsh} \cite{TullyMolecular1990, GranucciIncluding2010, ShuNonadiabatic2022} or \gls{kcsdm} \cite{ZhuCoherent2004, JasperNonBorn2006, ShuImplementation2020}.

\begin{figure}
    \centering
    \includegraphics[width=\columnwidth]{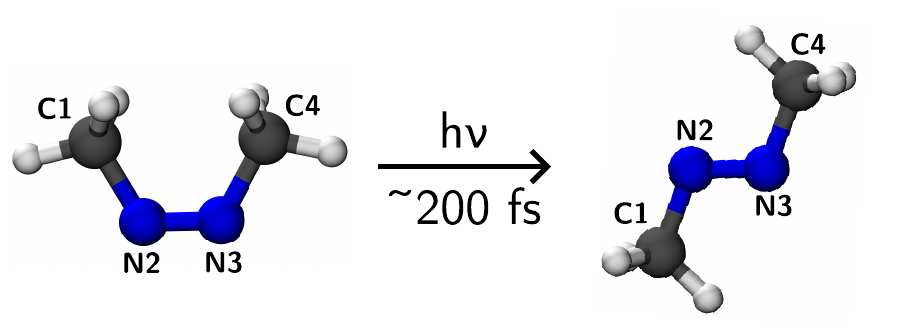}
    \caption{\label{fig:azomethane-photochemistry} The \textit{cis}-to-\textit{trans} photoisomerization of azomethane. blue: nitrogen, grey: carbon, white: hydrogen.}
\end{figure}

Here we implement a \SHARC-\PySCF interface that allows the \SHARC \cite{MaiNonadiabatic2018, MaiSHARC3.02023} dynamics program to use the \PySCF \cite{SunPySCF2018, SunRecent2020} electronic structure program for \gls{lpdft} \gls{ktsh} direct-dynamics calculations of the \textit{cis}-to-\textit{trans} photoisomerization reaction of azomethane (\cref{fig:azomethane-photochemistry}). One reason to choose this reaction for our initial study is that it has been widely studied both experimentally and theoretically \cite{MinezawaTrajectory2019, PapineauWhich2024, GhoshComparison2018, SellnerPhotodynamics2010, XuUltrafast2022, MerrittNonadiabatic2023, ZhaoNonadiabatic2023}. A second reason is the very informative recent study of this system by several methods \cite{PapineauWhich2024} that quantified how single-reference methods such as \gls{adc2} \cite{DreuwAlgebraic2015}, \gls{cc2} \cite{ChristiansenSecondorder1995}, \gls{cis} \cite{BeneSelf1971, ForesmanSystematic1992}, and \gls{tddft} \cite{CasidaTimeDependent1995, RungeDensityFunctional1984} are unable to satisfactorily simulate the \textit{cis}-to-\textit{trans} isomerization because the majority of the trajectories crash and that also found a large number of crashes when using multireference methods. Troublesome trajectories are a bane of this field because one must obtain continuous and smooth convergence of iterative calculations over the wide nuclear configuration space explored in typical photodissociation reactions. We study this problem in detail and show below that \gls{lpdft} is a robust method for this system, for which it yields results similar to \gls{xms-caspt2}\cite{GranovskyExtended2011}, while keeping the computational cost only slightly larger than that of the reference \gls{sa-casscf} method.

All electronic structure calculations reported here, except \gls{xms-caspt2} and frequency calculations, were performed with \PySCF\cite{SunPySCF2018, SunRecent2020} (Version 2.5.0, commit \texttt{v1.1-8184-geafc35752}) compiled with \libcint\cite{SunLibcint2015} (Version 6.1.1) and \libxc\cite{LehtolaRecent2018, Marqueslibxc2012} (Version 6.1.0) and used the \texttt{csf\_solver} from \mrh \cite{Hermesmrh2018}. Additionally, all \gls{lpdft} calculations used \PySCFforge\cite{PySCF2024} (commit \texttt{SHA-1 f817911}), an extension module for \PySCF. All geometry optimizations used the \geomeTRIC package\cite{WangGeometry2016} (Version 1.0) in \PySCF. All PDFT calculations used the tPBE on-top functional \cite{PerdewGeneralized1996, LiManniMulticonfiguration2014}. (All \gls{xms-caspt2} and \gls{sa-casscf} frequency calculations were performed in \OpenMolcas\cite{LiManniOpenMolcas2023} (Version 24.02, commit \texttt{v24.02-127-gd603295fc}). The \gls{xms-caspt2} calculations were performed with no \gls{ipea} shift \cite{GhigoModified2004}, with an imaginary level shift \cite{ForsbergMulticonfiguration1997} of \complexnum{0.3i}, and density fitting \cite{NishimotoAnalytic2022}. All software used here is free and open-source available to the whole community. 

No spatial symmetry was enforced in any calculation. All calculations of azomethane used the 6-31G* basis set \cite{HariharanInfluence1973} and state averaging over the lowest two singlet states (S$_0$ and S$_1$). We used a quadrature grid size of 4 (60/90 radial points and 434/590 angular points for atoms of periods 1 and 2 respectively. The \gls{lpdft} vertical excitation energies using grids 4 and 6 do not change (at least to the hundredths place). This agrees with our prior studies which showed that a level-3 grid is sufficient \cite{KingLarge2022, HennefarthLinearized2023a}.

We first investigate the suitability of \gls{lpdft} for describing the vertical excitation energy of the first excited singlet state of \textit{cis}- and \textit{trans}-azomethane, which is an $n\to\pi^*$ excitation. \Cref{tab:azomethane-vertical-excitation} summarizes the vertical excitation energies calculated using \gls{sa-casscf}, \gls{lpdft}, and \gls{xms-caspt2}; and, it also includes results from \gls{ri-adc2}, \gls{mrci} \cite{KnowlesInternally1992}, \gls{mrcisdq} \cite{WernerEfficient1988, KnowlesEfficient1988}, and \gls{cc3} \cite{ChristiansenSecondorder1995, KochCC31997}. The \gls{sa-casscf}, \gls{lpdft}, and \gls{xms-caspt2} vertical excitation energies were calculated at their equilibrium geometries optimized with their respective methods. \Cref{tab:azomethane-vertical-excitation} also includes experimental degassed aquesous \cite{HuttonPhotoisomerization1964} and gas phase \cite{RobinElectronic1967} \textit{trans}-isomer vertical excitation energies. The table shows that our results with two different active spaces agree well with the experimental results \cite{HuttonPhotoisomerization1964, RobinElectronic1967}, as do the \gls{xms-caspt2} and \gls{ri-adc2} calculations, and we find that the \gls{lpdft} calculations are more accurate than the \gls{sa-casscf}, \gls{mrci}, and \gls{mrcisdq} calculations. The good accuracy is consistent with our previous tests of \gls{lpdft} \cite{HennefarthLinearized2023, HennefarthLinearized2023a}.

\begin{table}
  \caption{\label{tab:azomethane-vertical-excitation} Vertical excitation energies (in \unit{\electronvolt}) of the first excited singlet state of azomethane at the \textit{cis} or \textit{trans} geometries.}
  \begin{tabular*}{\columnwidth}{@{\extracolsep{\fill}} l l c S[table-format=1.2] S[table-format=1.2]}
    \toprule\toprule
    Method & Basis & Active Space\footnote{The notation ($m$e,$n$o) denotes $m$ active electrons in $n$ active orbitals. The (6e,4o) active space is composed of the $\pi$ and $\pi^*$ orbitals and the two nitrogen lone-pair orbitals.} & {\textit{cis}} & {\textit{trans}} \\ 
    \midrule 
    \glsxtrshort{sa-casscf} & 6-31G* & (6e,4o) & 3.63 & 3.80 \\
    \glsxtrshort{sa-casscf} & 6-31G* & (10e,8o) & 3.85 & 3.99 \\
    \glsxtrshort{lpdft} & 6-31G* & (6e,4o) & 3.30 & 3.55 \\ 
    \glsxtrshort{lpdft} & 6-31G* & (10e,8o) & 3.31 & 3.52 \\ 
    \glsxtrshort{xms-caspt2} & 6-31G* & (6e,4o) & 3.38 & 3.60 \\
    \glsxtrshort{xms-caspt2} & 6-31G* & (10e,8o) & 3.47 & 3.68 \\
    \glsxtrshort{ri-adc2}\cite{SellnerPhotodynamics2010} & aug-cc-pVTZ & & 3.50 & 3.66 \\
    \glsxtrshort{mrci}\cite{SellnerPhotodynamics2010} & 6-31G* & (6e,4o) & 3.62 & 3.82 \\ 
    \glsxtrshort{mrcisdq}\footnote{Calculated at MP2/6-311G(2d,2p) optimized geometry. Reference orbitals are from \glsxtrshort{sa-casscf}(6e,4o) 
    calculation.}\cite{LiuAb1996} & cc-pVTZ & & & 3.77 \\
    \glsxtrshort{cc3}\footnote{Calculated at CCSD/cc-pVDZ optimized geometry.}\cite{SzalayTheoretical2011} & aug-cc-pVDZ & & & 3.76 \\
    expt. (aqueous) \cite{HuttonPhotoisomerization1964} & & & & 3.60 \\
    expt. (gas phase) \cite{RobinElectronic1967} & & & & 3.65 \\
    \bottomrule\bottomrule
  \end{tabular*}
\end{table}

Next we performed nonadiabatic molecular dynamics simulations. All dynamics simulations were performed using a locally modified version of \SHARC\cite{MaiNonadiabatic2018, MaiSHARC3.02023} (version 3.0, commit \texttt{SHA-1 afefdb8}) using \gls{ktsh} with energy-based decoherence. Specifically, we used the fewest switches algorithm \cite{TullyMolecular1990} with the curvature-driven approximation to the time-derivative coupling (calculated with the gradient formula \cite{ShuNonadiabatic2022}) with the energy-based decoherence scheme \cite{GranucciIncluding2010}. The curvature-driven approximation has been shown to perform similarly to calculating the full nonadiabatic coupling vector for this photoisomerization reaction \cite{ZhaoNonadiabatic2023}. An advantage of curvature-driven dynamics methods is that they do not require nonadiabatic coupling matrix elements or wave function overlaps between states with different geometries. A new interface script was added to allow \PySCF to be used as the direct-dynamics electronic structure solver. Initial conditions (geometries and velocities) were generated in \SHARC using a Wigner distribution without temperature broadening. \Gls{lpdft} and \gls{sa-casscf} trajectories were run with the same initial conditions where the harmonic frequencies used in the Wigner distribution were computed for the \gls{sa-casscf} ground state in \OpenMolcas. We used a fixed nuclear time step of \qty{0.5}{\fs}, and the electronic wave function was propagated using a \qty{0.0025}{\fs} time step. The decoherence parameter of the energy-based decoherence was set to \qty{0.1}{\hartrees}. The velocity vector was not modified when a frustrated hop was encountered. To conserve energy during a hop, the total velocity vector was rescaled, which is not likely to strongly impact the final dynamics \cite{BarbattiVelocity2021}. We performed dynamics calculations starting in the S$_1$ state of the \textit{cis} isomer and propagated each trajectory for \qty{150}{\fs} 

We first used the smaller (6e,4o) active space (fig. S1). We confirmed good energy conservation for each trajectory by verifying that the total energy variation was less than \qty{0.2}{\eV} for the entire trajectory and less than \qty{0.1}{\eV} between each step. We ran 150 trajectories with \gls{sa-casscf} and 150 with \gls{lpdft}, and only one \gls{sa-casscf} trajectory and five \gls{lpdft} trajectories failed to meet these energy conservation criteria (\cref{tab:num_failed}). The single problematic \gls{sa-casscf} trajectory very quickly failed to conserve energy (fig. S5) and also quickly failed for \gls{lpdft} (fig. S6); therefore we omit this trajectory from all further discussion because it is likely due to a bad initial condition. 
 
The \gls{lpdft} failure rate corresponds to only \qty{3}{\percent}, which compares favorably to a prior study of failure rates using \gls{xms-caspt2} with the same active space that reported over \qty{10}{\percent} of the trajectories crashing within \qty{150}{\fs} \cite{PapineauWhich2024} (although it was unstated which part of the calculations failed or if they could be restarted). We conclude that \gls{lpdft} \gls{ktsh} trajectories are less troublesome than trajectories with \gls{xms-caspt2} and that \gls{lpdft} is well-suited for dynamics simulations.

\begin{table}
    \caption{\label{tab:num_failed} Number of trajectories at each level of theory that did not conserve energy within \qty{0.2}{\eV}.}
    \centering
    \begin{tabular*}{\columnwidth}{@{\extracolsep{\fill}} l S[table-format=2] S[table-format=2]}
        \toprule\toprule 
        Method & {(6e,4o)} & {(10e,4o)} \\
        \midrule
        \glsxtrshort{sa-casscf} & 1  & 3 \\
        \glsxtrshort{lpdft}    & 5  & 7 \\ 
        \bottomrule\bottomrule
    \end{tabular*}
\end{table}

\begin{figure*}
    \centering
    \includegraphics[width=\textwidth]{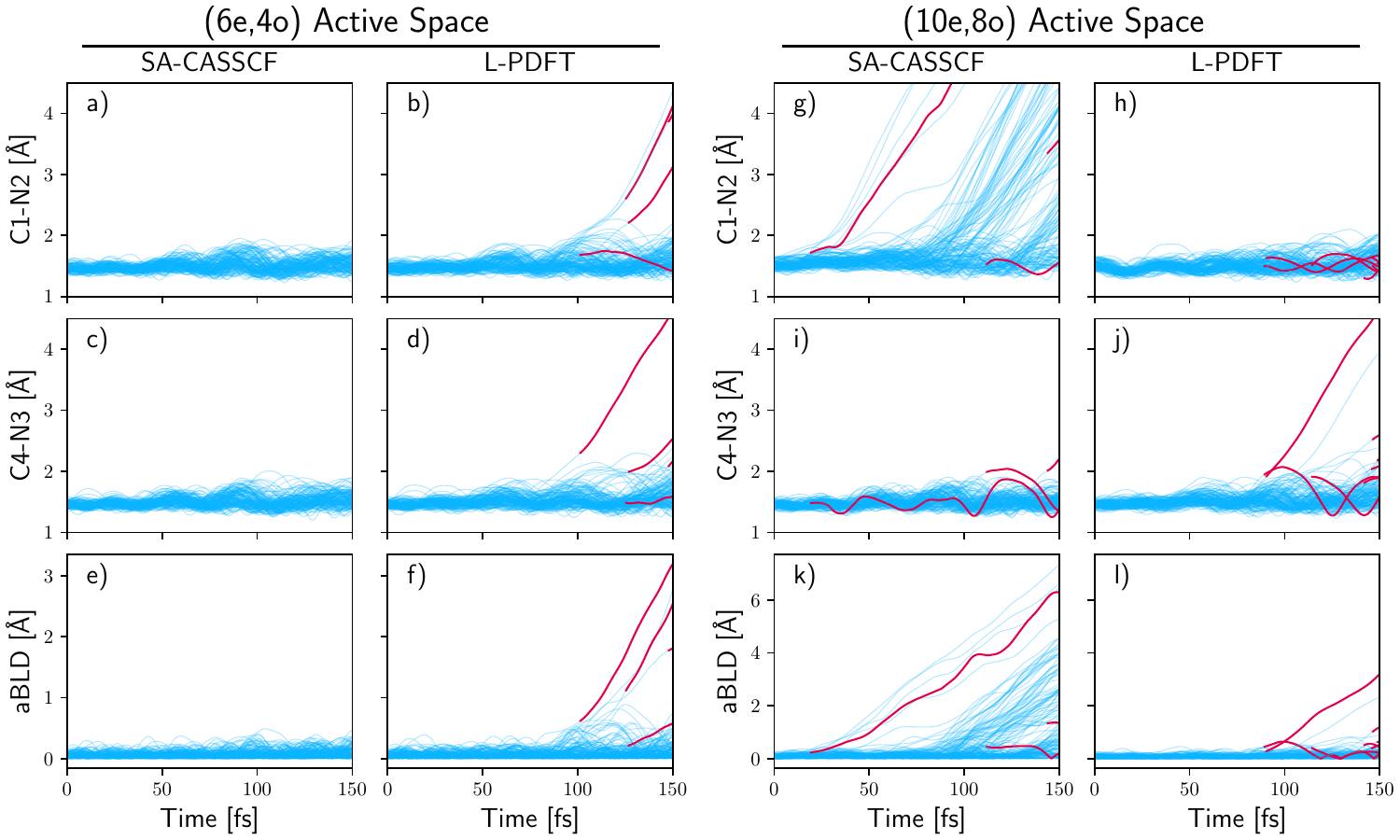}
    \caption{\label{fig:cis-bond} Time evolution of the C1-N2 and C4-N3 bonds as well as the \glsfirst{abld} between these two bonds for both the (6e,4o) and (10e,8o) active space. Blue curves represent trajectories that terminated normally and red curves represent portions of trajectories that no longer conserved total energy.}
\end{figure*}

In order to better understand the cause of trajectory failure, we inspected the \gls{lpdft} trajectories that did not conserve energy. We found that energy-conservation failure occurred after relaxation to the S$_0$ state and when one of the \ce{C-N} bonds is breaking. \Cref{fig:cis-bond} shows the \ce{C-N} bond lengths as functions of time as well as the \gls{abld} for both the \gls{sa-casscf} and \gls{lpdft} trajectories. Trajectories are colored in red after the point at which they no longer conserve energy. \Gls{sa-casscf} with the (6e,4o) active space does not have any trajectories for which the \ce{C-N} breaks, which is consistent with prior simulations using this active space \cite{XuUltrafast2022, SellnerPhotodynamics2010}. On the other hand, for each of the (6e,4o) \gls{lpdft} trajectories that does not conserve energy, the bond has already broken or is breaking. Since the (6e,4o) active space does not include any \ce{C-N} bonding or anti-bonding orbitals, it is likely that orbitals are rotating into and out of the active space in order to describe the bond breaking. This kind of behavior can cause discontinuous derivatives of the \gls{sa-casscf} potential energy surface, and it causes discontinuous potential energy surfaces for post-\gls{sa-casscf} methods such as \gls{lpdft} and \gls{xms-caspt2}. The lack of non-energy-conserving trajectories in the \gls{sa-casscf} calculations is because \gls{sa-casscf} never explores the dissociation region of the potential energy surface. 

\begin{figure*}
    \centering
    \includegraphics[width=0.9\textwidth]{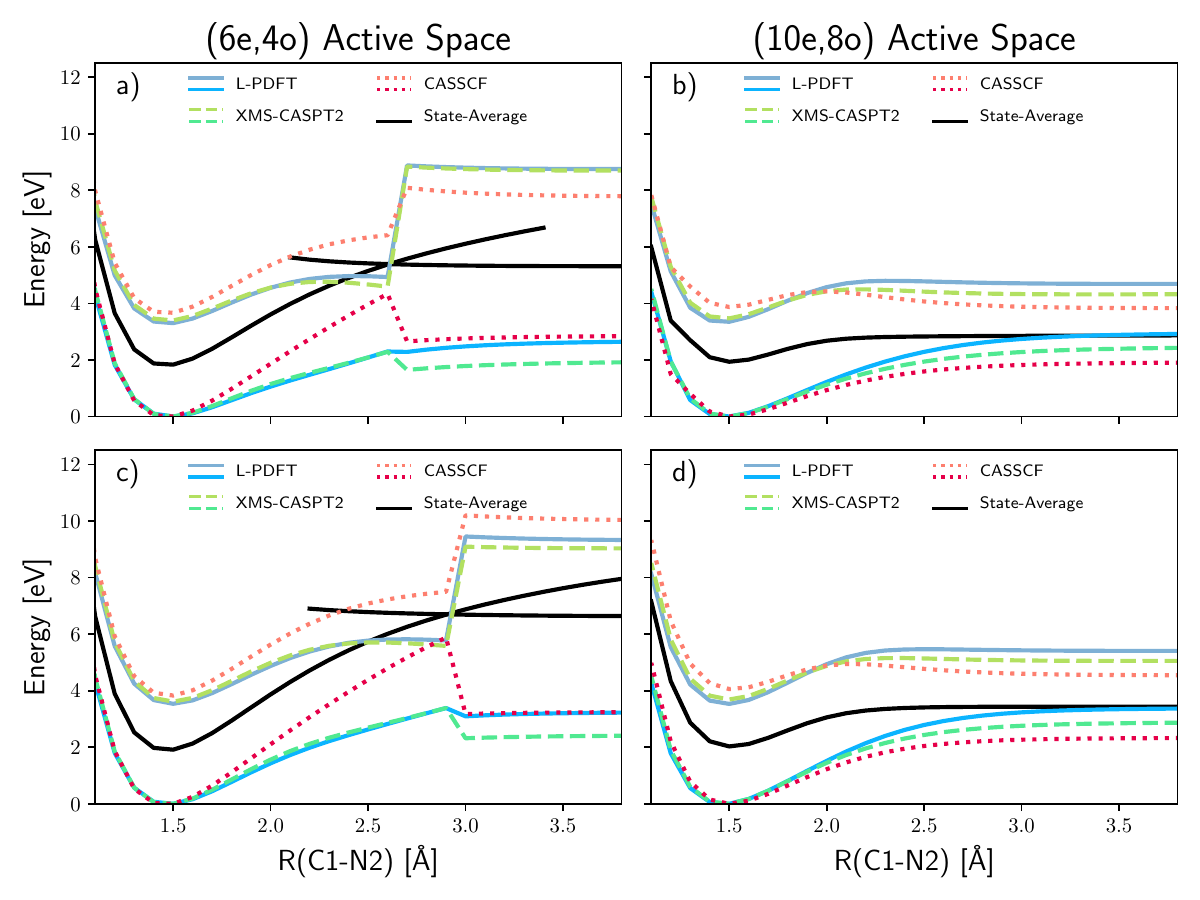}
    \caption{\label{fig:cn-scan} Potential energy scans of the lowest two singlet states of azomethane as computed with \gls{xms-caspt2}, \gls{lpdft}, and \gls{sa-casscf} for two directions of scanning ((a,b) starting with \textit{cis}; (c,d) starting with \textit{trans}), each with with two different active spaces ((a,c) using (6e,4o); (b,d) using (10e,8o)). Starting structures are from the \gls{lpdft} optimized structures at each respective isomer. Black curves represent the state-averaged \gls{sa-casscf} energy.}
\end{figure*}

To confirm the above interpretation of the problematic trajectories, we performed rigid scans of the energy along the C1-N2 bond-breaking coordinate starting from the \gls{lpdft} optimized \textit{cis}- and \textit{trans}-isomer structures. \Cref{fig:cn-scan} shows the potential energy curves for the S$_0$ and S$_1$ states computed by \gls{sa-casscf}, \gls{xms-caspt2}, and \gls{lpdft}. Starting with the orbitals in fig. S1, the \textit{cis} scan proceeds smoothly until about \qty{3.4}{\angstrom} at which point the solutions disappear. Scanning in the reverse direction shows a different solution, which can be followed until about \qty{2.2}{\angstrom}, with the energy of this solution crossing that of the forward scan solution. At the crossing point, the minimum-energy \gls{sa-casscf} solution must switch with a discontinuous derivative, and the crossing causes discontinuous curves for \gls{lpdft} and \gls{xms-caspt2}. This is unsurprising because the (6e,4o) active space is not appropriate for describing the \ce{C-N} photodissociation process. Panels a and c of \cref{fig:cn-scan} show that the \gls{sa-casscf} potential energy curves require a relatively large amount of energy for \ce{C-N} dissociation, even before the cross-over point (in particular about \qty{4}{\eV} and \qty{5}{\eV} on the ground state for the \textit{cis} and \textit{trans} isomers respectively in agreement with prior theoretical studies \cite{XuUltrafast2022}. \Gls{lpdft} agrees more closely with \gls{xms-caspt2} which has a substantially lower energy requirement for \ce{C-N} dissociation before the cross-over point, about \qty{2}{\eV}. 

Experimentally and theoretically, it is known that \ce{C-N} bond dissociation (after relaxation to the electronic ground state) is a possible photoproduct for azomethane \cite{XuUltrafast2022}. Prior simulations using \gls{xms-caspt2} with the same (6e,4o) active space \cite{XuUltrafast2022} and \gls{mrci} \cite{SellnerPhotodynamics2010}, and experiments \cite{DiauFemtosecond1998, DiauFemtochemistry2003} of \textit{cis}- and \textit{trans}-azomethane photoisomerization also had several \ce{C-N} bonds dissociating within \qty{150}{\fs}, although the computational studies did not specify how well any of these trajectories conserved total energy. Given the inability of the smaller (6e,4o) active space to properly describe this bond dissociation, it is likely that a significant number of \gls{xms-caspt2} trajectories in the prior study did not conserve energy after one of the \ce{C-N} bonds broke. Evidently, the failure of a few \gls{lpdft} trajectories to conserve energy with the (6e,4o) active space is due to the inability of this small active space to properly describe the \ce{C-N} dissociation, rather than being an issue with the \gls{lpdft} method. 

\begin{figure}
    \centering
    \includegraphics[width=\columnwidth]{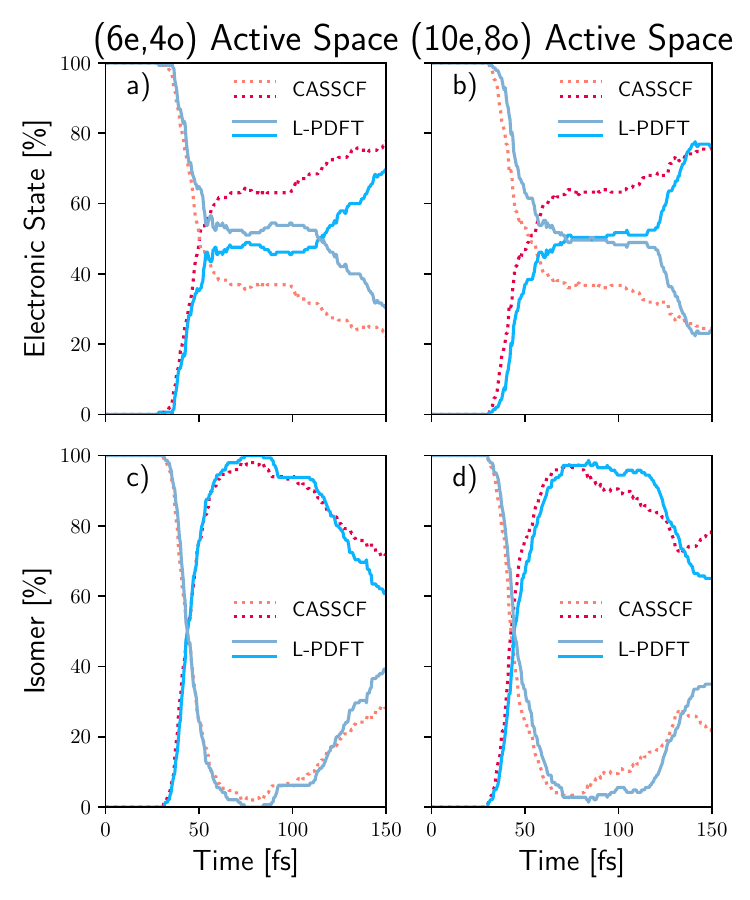}
    \caption{\label{fig:cis-pop-isomer} Ensemble-averaged electronic state population (a,b) and isomer percentages (c,d) as functions of time using the smaller (6e,4o) (a,c) and larger (10e,8o) (b,d) active spaces. All trajectories were started in the S$_1$ state of the \textit{cis} isomer.}
\end{figure}

To model the \ce{C-N} bond dissociation pathway, we carried out dynamics calculations using the larger (10e,8o) active space that includes the C1-N2 bonding and antibonding orbitals (fig. S3 and S4). The (10e,8o) active space also includes N-N $\sigma$ and $\sigma^*$ orbitals. Although our initial active space is not symmetric, the molecular geometry during the course of the simulation will rarely (if ever) have any point group symmetry. Using the larger active space, all of the trajectories finished, and only three \gls{sa-casscf} and seven \gls{lpdft} trajectories of the 150 total trajectories (less than \qty{5}{\percent}) failed to conserve total energy (\cref{tab:num_failed}). Again this compares favorably to the previous \gls{xms-caspt2} study, which had over \qty{10}{\percent} crashed trajectories. Inspection of the \ce{C-N} bonds in the new batch of trajectories shows that \gls{sa-casscf} now has many trajectories with C1-N2 bond breaking and no trajectories with C4-N3 bond breaking. This is rationalized from the rigid potential energy curves in panels b and d of \cref{fig:cn-scan} where \gls{sa-casscf} with the (10e,8o) active space predicts a lower energy for \ce{C-N} dissociation as compared to the smaller (6e,4o) active space. However, since there are no C4-N3 bonding/antibonding orbitals in the active space, it is likely that there needs to be a significant excess of energy, similar to that seen in the (6e,4o) active space, to dissociate the C4-N3 bond. Further, the number of trajectories exhibiting \ce{C-N} bond dissociation, and the fast speed at which dissociation occurs (bonds breaking almost immediately), are unphysical. A prior study of \gls{sa-casscf} dynamics using \textit{ab initio} multiple spawning and a similar (10e,8o) active space also had a significant number of trajectories with \ce{C-N} bond dissociation \cite{GhoshComparison2018}. However, experimental evidence suggests that the photodissociation is a minor product and that bond dissociation does not occur on the excited state and should occur only after the molecule has reached the ground state (no trajectories reach the ground state until after \qty{30}{\fs} as shown in \cref{fig:cis-pop-isomer}) \cite{DiauFemtosecond1998, DiauFemtochemistry2003, NorthUltraviolet1993, BrackerUltraviolet1998}. \Gls{lpdft} no longer has any C1-N2 bond dissociations, and instead only the C4-N3 bonds are breaking (\cref{fig:cn-scan}). This is likely due to the active space not containing the C4-N3 bonding or antibonding orbitals giving results similar to the (6e,4o) active space. When energy nonconservation occurs with the larger active space, it is likely due to the inadequacy of this active space to properly model the C4-N3 bond breaking. An even larger active space would be required in order to properly model both \ce{C-N} bond dissociations; although this is computationally feasible for this small system, it is beyond the scope of the current letter.

\begin{figure}
    \centering
    \includegraphics[width=\columnwidth]{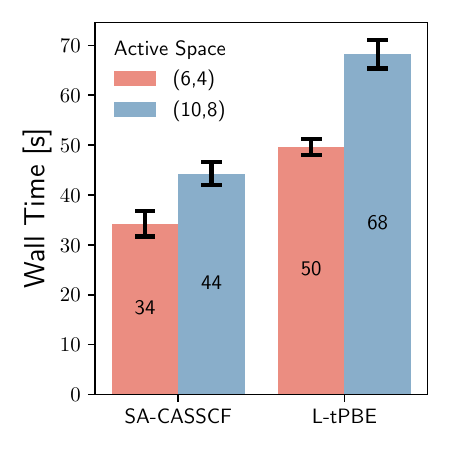}
    \caption{\label{fig:timings} Average wall time for each step of the dynamics. Only normal terminating trajectories are considered. Black bars represent standard deviation. }
\end{figure}

Using only those trajectories that conserved energy and terminated normally, \cref{fig:cis-pop-isomer} shows the time-dependent electronic state populations and isomer percentage for each method and active space. Both \gls{sa-casscf} and \gls{lpdft} have an induction period of about \qty{30}{\fs} and then rapidly decay from S$_1$ to S$_0$ until about \qty{60}{\fs}. During this time, almost all trajectories have switched to the \textit{trans} isomer. A second plateau in the electronic populations then lasts for about \qty{45}{\fs} at which point the molecule reaches a second strong-coupling region as some molecules go from \textit{trans} to \textit{cis} isomers. The primary difference between the two methods is the speed at which the molecules relax to the ground state. \Gls{sa-casscf} predicts faster decay with over \qty{60}{\percent} of the trajectories relaxing to the S$_0$ state after passing the first region of strong-coupling whereas \gls{lpdft} predicts about \qty{45}{\percent}. Prior studies have similarly noted that \gls{sa-casscf} overestimates the short-time decay rate as compared to both \gls{xms-caspt2} and \gls{ms-caspt2} \cite{MerrittNonadiabatic2023, PapineauWhich2024, XuUltrafast2022, ZhaoNonadiabatic2023}. 

\Cref{fig:cis-pop-isomer} of the present study may be directly compared to fig. 1 of \citet{PapineauWhich2024}. This shows that the present \gls{lpdft} ensemble-averaged populations and isomer percentages, for both active spaces, agree well the prior \gls{ms-caspt2} results. This shows that \gls{lpdft} is a robust method that can yield results as accurate as the more expensive \gls{ms-caspt2} and \gls{xms-caspt2} method.

We also confirm that the \gls{lpdft} simulations may be carried out efficiently. \Cref{fig:timings} summarizes the average wall time for each step of the dynamics for both \gls{sa-casscf} and \gls{lpdft} with both active spaces. All calculations were performed using 2 cores with \qty{8}{\giga\byte} of memory on an Intel Xeon Gold 6248R processor. Each step involves a single-point calculation as well as two gradient calculations: one for each state. The gradient calculations were done in parallel, with each using a single core. The wall time for each step represents the time for the single-point calculation and the maximum time needed to perform either of the gradient calculations plus any other time needed by the \SHARC driver (which corresponds to ``overhead'' and should be minimal). For both active spaces, \gls{lpdft} steps take on average only 1.5 times as long as \gls{sa-casscf}.

 Our implementation so far is not fully optimized. For example, caching intermediate quantities within the \SHARC-\PySCF interface could speed up gradient calculations, and implementation of density fitting \cite{ScottAnalytic2021} could also be used to speed up future calculations.

In this letter, we have presented the implementation of an interface between \SHARC and \PySCF in order to leverage the accuracy and computational efficiency of \gls{lpdft} for nonadiabatic molecular dynamics. To demonstrate the capability of the method, we have have studied the \textit{cis}-to-\textit{trans} photoisomerization of azomethane. Utilizing both a smaller and a larger active space, no \gls{lpdft} trajectories crashed, which compares favorably with prior studies utilizing \gls{xms-caspt2} for the same system where more than \qty{10}{\percent} of the trajectories crashed \cite{PapineauWhich2024}. Like \gls{xms-caspt2}, \gls{lpdft} highlights the importance of the \ce{C-N} photodissociation product. We found that only a small fraction of the trajectories did not conserve energy (\cref{tab:num_failed}), and we showed that the few failures found were due to the active space not properly modeling \ce{C-N} dissociation. We showed that \gls{lpdft} adds only a small computational cost over \gls{sa-casscf} (\cref{fig:timings}), and it provides a slower population decay more similar to \gls{ms-caspt2} \cite{PapineauWhich2024}. In summary, we find that \gls{lpdft} is a promising new multireference method that is able to provide results similar to the much more expensive multireference perturbation theories at a computational cost only slightly more expensive than \gls{sa-casscf}. We also found that \gls{lpdft} is more robust than traditional multireference perturbation theories for molecular dynamics. 

\section*{Supporting Information}
Active spaces used, total energy deviation as a function of time, geometrical distributions at surface hops, optimized azomethane coordinates, harmonic frequencies used in the Wigner distribution, and example input files for \SHARC.

\begin{acknowledgments}
  This work was supported in part by the Air Force Office Scientific Research
  (grant no. FA9550-20-1-0360). M.R.H. acknowledges support by the National
  Science Foundation Graduate Research Fellowship Program (grant no. 2140001).
  Any opinions, findings, and conclusions or recommendations expressed in this
  material are those of the author(s) and do not necessarily reflect the views
  of the National Science Foundation. We also acknowledge the University of
  Chicago’s Research Computing Center for their support of this work.
\end{acknowledgments}

\bibliographystyle{achemso}
\bibliography{ms}

\section*{TOC Graphic}
\includegraphics[width=\columnwidth]{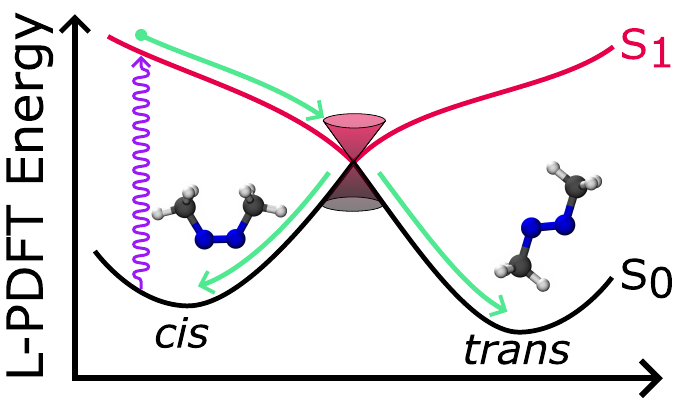}

\end{document}

\title{Supporting Information: Semiclassical Nonadiabatic Molecular Dynamics Using Linearized Pair-Density Functional Theory}

\author{Matthew R. Hennefarth}
\affiliation{Department of Chemistry and Chicago Center for Theoretical Chemistry, University of Chicago, Chicago, IL 60637, USA}

\author{Donald G. Truhlar}
 \email[corresponding author: ]{truhlar@umn.edu}
\affiliation{Department of Chemistry, Chemical Theory Center, and Minnesota Supercomputing Institute, University of Minnesota, Minneapolis, MN 55455-0431, USA}

\author{Laura Gagliardi}
 \email[corresponding author: ]{lgagliardi@uchicago.edu}
\affiliation{Department of Chemistry, Pritzker School of Molecular Engineering, The James Franck Institute, and Chicago Center for Theoretical Chemistry, University of Chicago, Chicago, IL 60637, USA}
\affiliation{Argonne National Laboratory, 9700 S. Cass Avenue, Lemont, IL 60439, USA}

%\date{\today}
\date{September 13, 2024}

\maketitle

\tableofcontents

\section{Supplemental Figures}
\begin{figure}[H]
    \centering
    \includegraphics[width=4.5in]{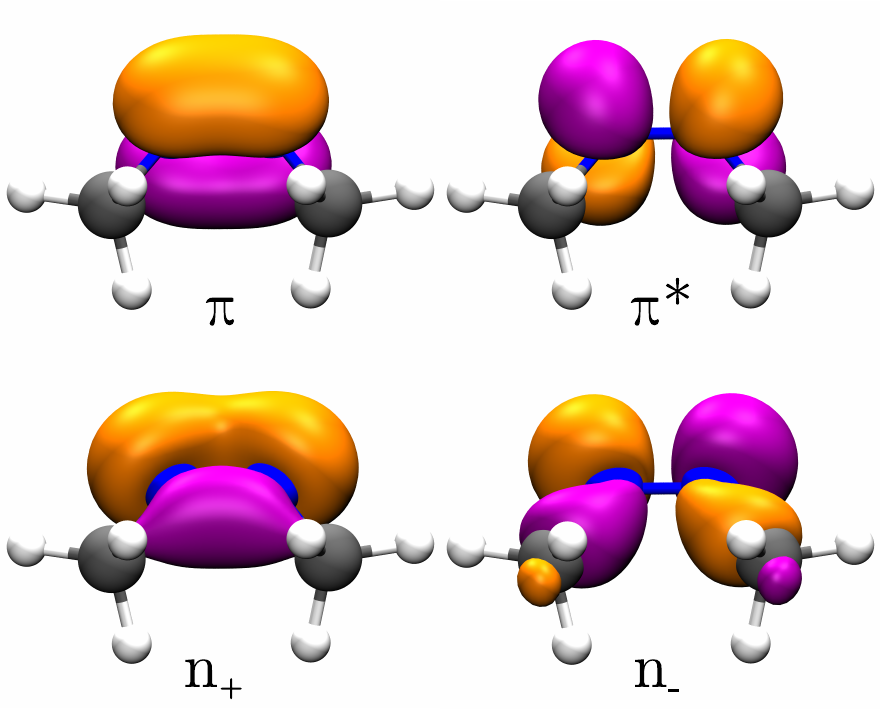}
    \caption{Smaller (6,4) active space natural orbitals used for the \textit{cis}-azomethane calculations consisting of the \ce{N-N} $\pi,\pi^*$ orbitals and  the symmetric ($n^+$) and antisymmetric ($n^-$) nitrogen lone pair orbitals. The geometry is optimized at the L(2)-tPBE(6,4)/6-31G* level of theory. Isosurface value of 0.06.}
\end{figure}

\begin{figure}[H]
    \centering
    \includegraphics[width=4.5in]{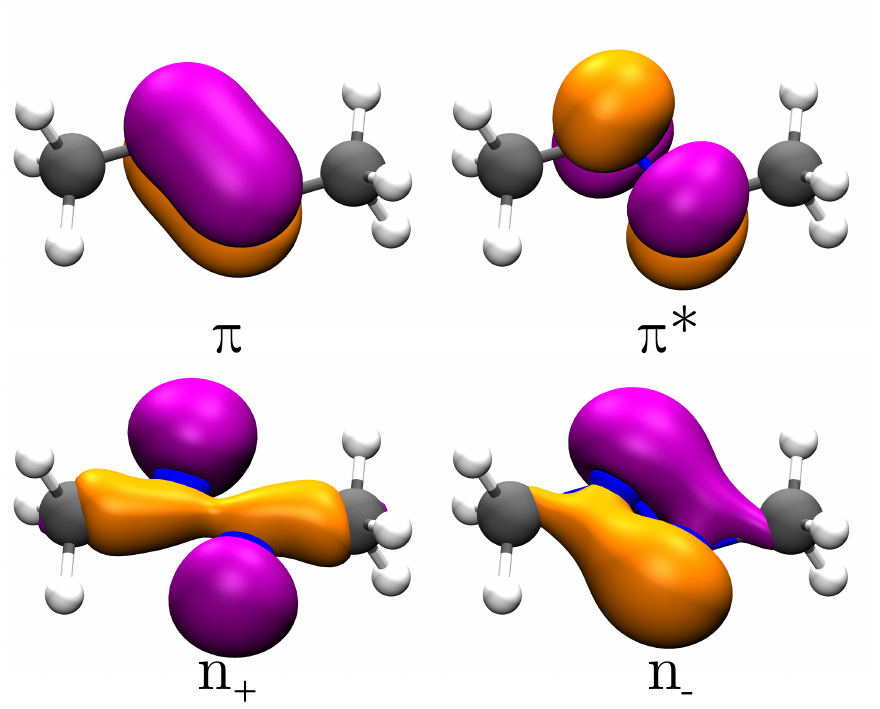}
    \caption{Smaller (6,4) active space natural orbitals used for the \textit{trans}-azomethane calculations consisting of the \ce{N-N} $\pi,\pi^*$ orbitals and  the symmetric ($n^+$) and antisymmetric ($n^-$) nitrogen lone pair orbitals. The geometry is optimized at the L(2)-tPBE(6,4)/6-31G* level of theory. Isosurface value of 0.06.}
\end{figure}

\begin{figure}[H]
    \centering
    \includegraphics[width=7in]{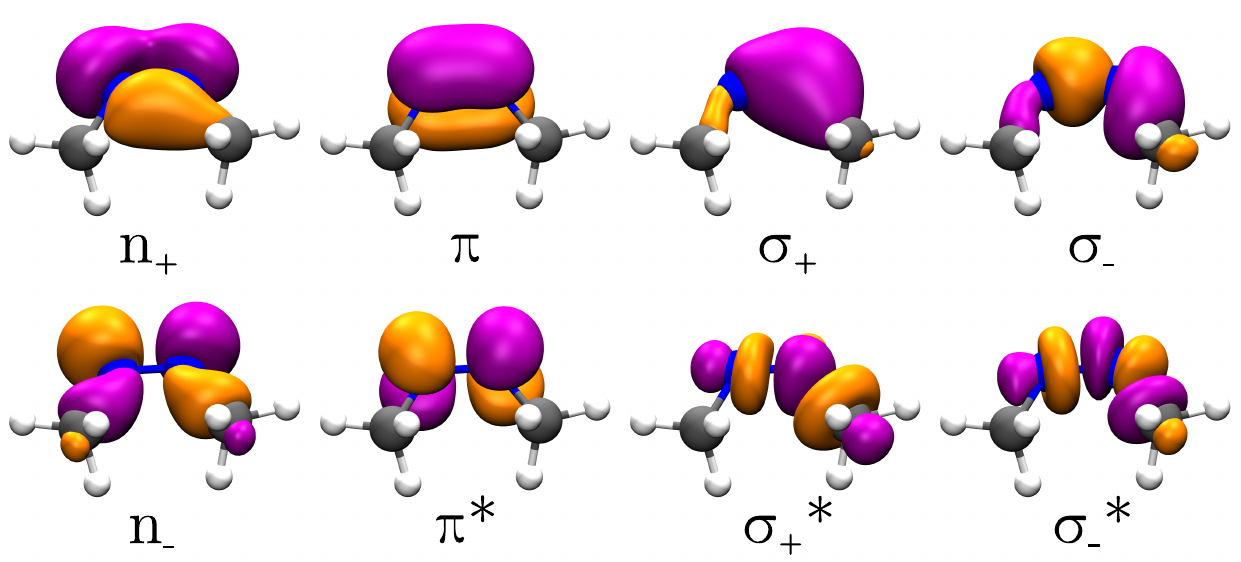}
    \caption{Larger (10, 8) active space natural orbitals used for the \textit{cis}-azomethane calculations consisting of the \ce{N-N} $\pi,\pi^*$ orbitals, the symmetric ($n^+$) and antisymmetric ($n^-$) nitrogen lone pair orbitals, and symmetric ($\sigma_+,\sigma_+^*$) and antisymmetric ($\sigma_-,\sigma_-^*$) \ce{N-N} and \ce{N-C} bonding and anti-bonding orbitals. The geometry is optimized at the L(2)-tPBE(10,8)/6-31G* level of theory. Isosurface value of 0.06.}
\end{figure}

\begin{figure}[H]
    \centering
    \includegraphics[width=7in]{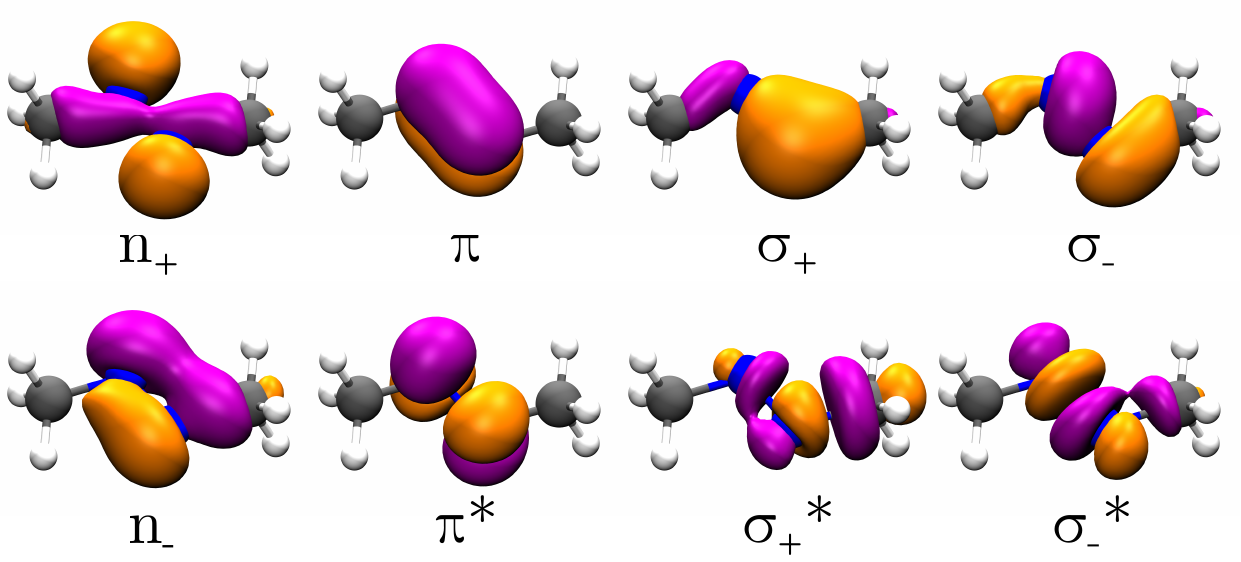}
    \caption{Larger (10, 8) active space natural orbitals used for the \textit{trans}-azomethane calculations consisting of the \ce{N-N} $\pi,\pi^*$ orbitals, the symmetric ($n^+$) and antisymmetric ($n^-$) nitrogen lone pair orbitals, and symmetric ($\sigma_+,\sigma_+^*$) and antisymmetric ($\sigma_-,\sigma_-^*$) \ce{N-N} and \ce{N-C} bonding and anti-bonding orbitals. The geometry is optimized at the L(2)-tPBE(10,8)/6-31G* level of theory. Isosurface value of 0.06.}
\end{figure}

\begin{figure}[H]
    \centering
    \includegraphics{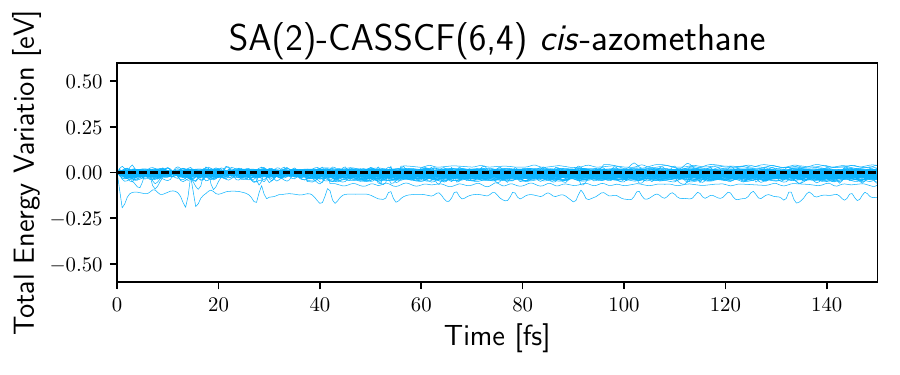}
    \caption{Total energy variation as functions of time for SA(2)-CASSCF(6,4) starting in the \textit{cis} isomer.}
\end{figure}

\begin{figure}[H]
    \centering
    \includegraphics{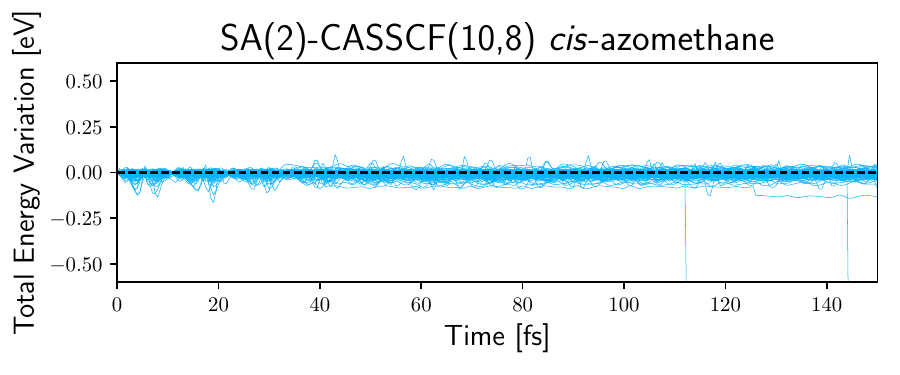}
    \caption{Total energy variation as functions of time for SA(2)-CASSCF(10,8) starting in the \textit{cis} isomer.}
\end{figure}

\begin{figure}[H]
    \centering
    \includegraphics{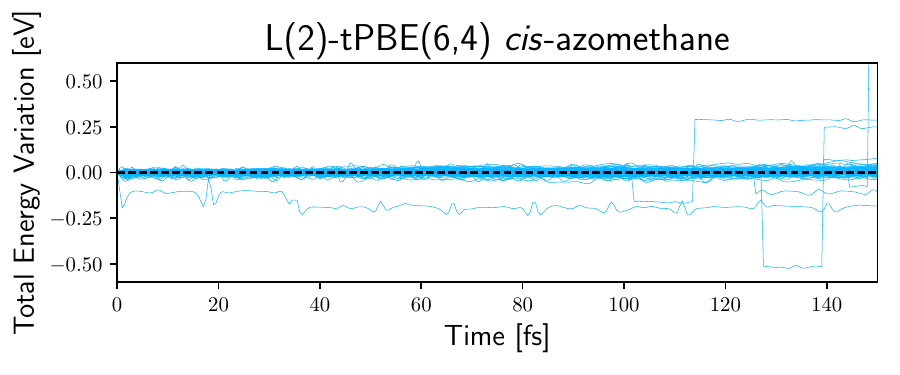}
    \caption{Total energy variation as functions of time for L(2)-tPBE(6,4) starting in the \textit{cis} isomer.}
\end{figure}

\begin{figure}[H]
    \centering
    \includegraphics{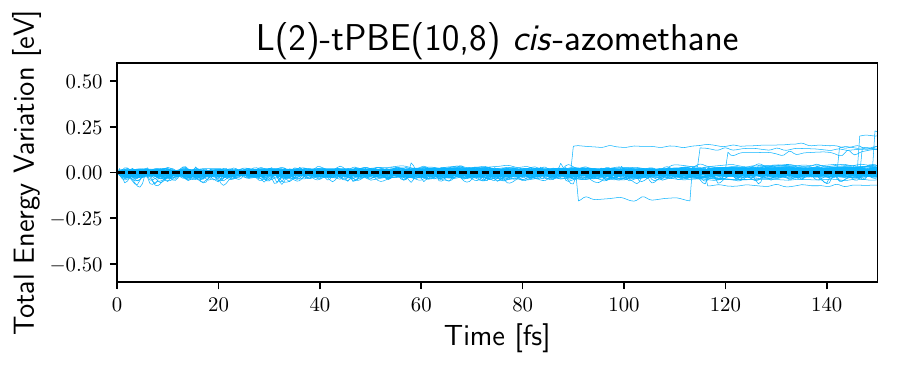}
    \caption{Total energy variation as functions of time for L(2)-tPBE(10,8) starting in the \textit{cis} isomer.}
\end{figure}

\section{Optimized Geometries}

\subsection{\textit{cis}-azomethane}

\begin{table}[H]
\centering
\caption{SA(2)-CASSCF(6,4)/6-31G* \textit{cis}-azomethane equilibrium ground state geometry (in \unit{\angstrom}). State 0 energy: \qty{-188.08357639155}{\hartees}. State 1 energy: \qty{-187.950097843184}{\hartees}}
\begin{tabular}{l d d d}
N & 0.10301 & -0.00000 & -0.02243 \\
N & 0.10302 & 0.00000 & 1.22244 \\
C & 1.36738 & -0.00000 & 1.95857 \\
C & 1.36737 & 0.00000 & -0.75857 \\
H & 1.12869 & -0.00000 & -1.81190 \\
H & 1.95786 & 0.88156 & -0.53395 \\
H & 1.95786 & -0.88156 & -0.53395 \\
H & 1.12870 & 0.00000 & 3.01190 \\
H & 1.95787 & -0.88156 & 1.73395 \\
H & 1.95787 & 0.88156 & 1.73395 \\
\end{tabular}
\end{table}

\begin{table}[H]
\centering
\caption{SA(2)-CASSCF(10,8)/6-31G* \textit{trans}-azomethane equilibrium ground state geometry (in \unit{\angstrom}). State 0 energy: \qty{-188.151684469685}{\hartees}. State 1 energy: \qty{-188.010337616342}{\hartees}}
\begin{tabular}{l d d d}
N & 0.07074 & -0.00000 & -0.01687 \\
N & 0.09137 & -0.00000 & 1.24884 \\
C & 1.36379 & 0.00000 & 1.96564 \\
C & 1.38946 & -0.00000 & -0.77415 \\
H & 1.13468 & -0.00000 & -1.82306 \\
H & 1.96992 & 0.88500 & -0.54378 \\
H & 1.96992 & -0.88501 & -0.54378 \\
H & 1.14078 & 0.00000 & 3.02257 \\
H & 1.94948 & -0.88187 & 1.73231 \\
H & 1.94948 & 0.88187 & 1.73231 \\
\end{tabular}
\end{table}

\begin{table}[H]
\centering
\caption{L(2)-tPBE(6,4)/6-31G* \textit{cis}-azomethane equilibrium ground state geometry (in \unit{\angstrom}). State 0 energy: \qty{-188.981625763722}{\hartees}. State 1 energy: \qty{-188.860488885797}{\hartees}}
\begin{tabular}{l d d d}
N & 0.07770 & 0.00000 & -0.02126 \\
N & 0.07770 & -0.00000 & 1.22126 \\
C & 1.36730 & 0.00000 & 1.96012 \\
C & 1.36730 & -0.00000 & -0.76013 \\
H & 1.13090 & 0.00000 & -1.83272 \\
H & 1.96945 & 0.89576 & -0.52359 \\
H & 1.96945 & -0.89576 & -0.52359 \\
H & 1.13090 & -0.00000 & 3.03271 \\
H & 1.96945 & -0.89576 & 1.72358 \\
H & 1.96945 & 0.89576 & 1.72358 \\
\end{tabular}
\end{table}

\begin{table}[H]
\centering
\caption{L(2)-tPBE(10,8)/6-31G* \textit{cis}-azomethane equilibrium ground state geometry (in \unit{\angstrom}). State 0 energy: \qty{-188.97255638239}{\hartees}. State 1 energy: \qty{-188.851010312274}{\hartees}}
\begin{tabular}{l d d d}
N & 0.07872 & 0.00000 & -0.02725 \\
N & 0.07846 & -0.00000 & 1.22145 \\
C & 1.36691 & 0.00000 & 1.96144 \\
C & 1.36558 & -0.00000 & -0.76078 \\
H & 1.13437 & 0.00000 & -1.83453 \\
H & 1.96817 & 0.89509 & -0.52342 \\
H & 1.96817 & -0.89509 & -0.52342 \\
H & 1.12979 & -0.00000 & 3.03391 \\
H & 1.96971 & -0.89577 & 1.72628 \\
H & 1.96971 & 0.89577 & 1.72628 \\
\end{tabular}
\end{table}

\begin{table}[H]
\centering
\caption{XMS(2)-CASPT2(6,4)/6-31G* \textit{cis}-azomethane equilibrium ground state geometry (in \unit{\angstrom}). State 0 energy: \qty{-188.64165234}{\hartees}. State 1 energy: \qty{-188.51734293}{\hartees}}
\begin{tabular}{l d d d}
N & 0.08101 & -0.00000 & -0.02854 \\
N & 0.08102 &  0.00000 &  1.22853 \\
C & 1.37150 & -0.00000 &  1.95678 \\
C & 1.37150 &  0.00000 & -0.75679 \\
H & 1.13692 & -0.00000 & -1.82053 \\
H & 1.96267 &  0.89026 & -0.51646 \\
H & 1.96267 & -0.89026 & -0.51646 \\
H & 1.13692 &  0.00000 &  3.02052 \\
H & 1.96267 & -0.89026 &  1.71645 \\
H & 1.96267 &  0.89026 &  1.71645 \\
\end{tabular}
\end{table}

\begin{table}[H]
\centering
\caption{XMS(2)-CASPT2(10,8)/6-31G* \textit{cis}-azomethane equilibrium ground state geometry (in \unit{\angstrom}). State 0 energy: \qty{-188.64236168}{\hartees}. State 1 energy: \qty{-188.51489981}{\hartees}}
\begin{tabular}{l d d d}
N & 0.07662636 & -0.00000058 & -0.02808695 \\
N & 0.08123588 & 0.00000056 & 1.23096375 \\
C & 1.37101405 & -0.00000013 & 1.95816387 \\
C & 1.37358989 & 0.00000015 & -0.75946656 \\
H & 1.13807799 & -0.00000009 & -1.82268836 \\
H & 1.96346666 & 0.89019838 & -0.51909243 \\
H & 1.96346693 & -0.89019787 & -0.51909223 \\
H & 1.13772167 & 0.00000008 & 3.02223585 \\
H & 1.96220512 & -0.89029263 & 1.71850668 \\
H & 1.96220543 & 0.89029212 & 1.71850640 \\
\end{tabular}
\end{table}

\subsection{\textit{trans}-azomethane}

\begin{table}[H]
\centering
\caption{SA(2)-CASSCF(6,4)/6-31G* \textit{trans}-azomethane equilibrium ground state geometry (in \unit{\angstrom}). State 0 energy: \qty{-188.103159734455}{\hartees}. State 1 energy: \qty{-187.963485191687}{\hartees}}
\begin{tabular}{l d d d}
C & 1.42953 & -0.00007 & 1.67795 \\
N & 0.05054 & -0.00036 & 1.21666 \\
N & -0.04952 & 0.00221 & -0.01951 \\
C & -1.42851 & 0.00182 & -0.48081 \\
H & -2.13503 & -0.00070 & 0.34027 \\
H & -1.57510 & -0.87250 & -1.10414 \\
H & -1.57675 & 0.87865 & -1.10017 \\
H & 2.13605 & 0.00249 & 0.85687 \\
H & 1.57786 & -0.87688 & 2.29736 \\
H & 1.57603 & 0.87428 & 2.30124 \\
\end{tabular}
\end{table}

\begin{table}[H]
\centering
\caption{SA(2)-CASSCF(10,8)/6-31G* \textit{trans}-azomethane equilibrium ground state geometry (in \unit{\angstrom}). State 0 energy: \qty{-188.170501871954}{\hartees}. State 1 energy: \qty{-188.024022048082}{\hartees}}
\begin{tabular}{l d d d}
C & 1.45314 & -0.00007 & 1.69160 \\
N & 0.02402 & -0.00045 & 1.23536 \\
N & -0.05818 & 0.00224 & -0.02640 \\
C & -1.43702 & 0.00185 & -0.49026 \\
H & -2.14743 & -0.00071 & 0.32701 \\
H & -1.58171 & -0.87246 & -1.11389 \\
H & -1.58348 & 0.87866 & -1.10994 \\
H & 2.14130 & 0.00246 & 0.85704 \\
H & 1.59811 & -0.88018 & 2.30565 \\
H & 1.59636 & 0.87760 & 2.30954 \\
\end{tabular}
\end{table}

\begin{table}[H]
\centering
\caption{L(2)-tPBE(6,4)/6-31G* \textit{trans}-azomethane equilibrium ground state geometry (in \unit{\angstrom}). State 0 energy: \qty{-188.99634453752}{\hartees}. State 1 energy: \qty{-188.865879862179}{\hartees}}
\begin{tabular}{l d d d}
C & 1.43882 & 0.00006 & 1.67998 \\
N & 0.04077 & -0.00162 & 1.21943 \\
N & -0.03974 & 0.00091 & -0.02229 \\
C & -1.43780 & 0.00190 & -0.48283 \\
H & -2.15752 & 0.00003 & 0.35275 \\
H & -1.58278 & -0.88537 & -1.12157 \\
H & -1.58291 & 0.89195 & -1.11768 \\
H & 2.15854 & 0.00338 & 0.84440 \\
H & 1.58555 & -0.88981 & 2.31469 \\
H & 1.58218 & 0.88750 & 2.31885 \\
\end{tabular}
\end{table}

\begin{table}[H]
\centering
\caption{L(2)-tPBE(10,8)/6-31G* \textit{trans}-azomethane equilibrium ground state geometry (in \unit{\angstrom}). State 0 energy: \qty{-188.986328119442}{\hartees}. State 1 energy: \qty{-188.85690872463}{\hartees}}
\begin{tabular}{l d d d}
C & 1.44642 & -0.00045 & 1.67018 \\
N & 0.04836 & -0.00026 & 1.22108 \\
N & -0.04490 & 0.00387 & -0.02470 \\
C & -1.44689 & 0.00212 & -0.47350 \\
H & -2.15885 & -0.00118 & 0.36867 \\
H & -1.59699 & -0.88549 & -1.11060 \\
H & -1.60034 & 0.89193 & -1.10650 \\
H & 2.16068 & 0.00189 & 0.82986 \\
H & 1.59943 & -0.89004 & 2.30373 \\
H & 1.59821 & 0.88650 & 2.30750 \\
\end{tabular}
\end{table}

\begin{table}[H]
\centering
\caption{XMS(2)-CASPT2(6,4)/6-31G* \textit{trans}-azomethane equilibrium ground state geometry (in \unit{\angstrom}). State 0 energy: \qty{-188.65756616}{\hartees}. State 1 energy: \qty{-188.52500882}{\hartees}}
\begin{tabular}{l d d d}
C &  1.43442 & -0.00004 &  1.68125 \\
N &  0.03796 & -0.00042 &  1.22666 \\
N & -0.03694 &  0.00222 & -0.02951 \\
C & -1.43340 &  0.00182 & -0.48410 \\
H & -2.13676 & -0.00079 &  0.35317 \\
H & -1.57729 & -0.88082 & -1.11323 \\
H & -1.57910 &  0.88702 & -1.10921 \\
H &  2.13778 &  0.00249 &  0.84397 \\
H &  1.58009 & -0.88520 &  2.30642 \\
H &  1.57834 &  0.88264 &  2.31031 \\
\end{tabular}
\end{table}

\begin{table}[H]
\centering
\caption{XMS(2)-CASPT2(10,8)/6-31G* \textit{trans}-azomethane equilibrium ground state geometry (in \unit{\angstrom}). State 0 energy: \qty{-188.65712485}{\hartees}. State 1 energy: \qty{-188.52199685}{\hartees}}
\begin{tabular}{l d d d}
C &  1.43734 & -0.00000 &  1.68324 \\
N &  0.03468 & -0.00086 &  1.22847 \\
N & -0.03867 &  0.00182 & -0.02814 \\
C & -1.43449 &  0.00187 & -0.48550 \\
H & -2.13979 & -0.00054 &  0.35012 \\
H & -1.57828 & -0.88068 & -1.11458 \\
H & -1.57960 &  0.88701 & -1.11062 \\
H &  2.13814 &  0.00272 &  0.84457 \\
H &  1.58402 & -0.88533 &  2.30712 \\
H &  1.58175 &  0.88291 &  2.31104 \\
\end{tabular}
\end{table}

\section{Wigner Distribution Frequencies}

\begin{table}[H]
\centering
\caption{Harmonic frequencies (in \unit{\cm^{-1}}) used to generate the Wigner distribution. All values are calculated at the SA(2)-CASSCF/6-31G* level of theory.}
\begin{tabular*}{\columnwidth}{@{\extracolsep{\fill}}S[table-format=2] S[table-format=4.2] S[table-format=4.2]}
\toprule\toprule 
{Mode} & {(6e,4o)} & {(10e,8o)} \\ 
\midrule 
1 & 38.93 &  23.83 \\ 
2 & 219.43 &  218.74 \\ 
3 & 394.43 &  382.38 \\ 
4 & 483.23 &  462.43 \\ 
5 & 674.51 &  633.82 \\ 
6 & 988.19 & 868.79 \\ 
7 & 1079.67   & 1046.01 \\ 
8 & 1109.45   & 1060.56 \\ 
9 & 1207.55   & 1181.23 \\ 
10 & 1218.56  & 1187.10 \\ 
11 & 1310.13  & 1283.49 \\ 
12 & 1563.25  & 1529.18 \\ 
13 & 1581.86  & 1575.65 \\ 
14 & 1619.49  & 1592.23 \\ 
15 & 1629.09  & 1614.31 \\ 
16 & 1643.78  & 1638.29 \\ 
17 & 1666.49  & 1646.03 \\ 
18 & 1741.98  & 1663.33 \\ 
19 & 3213.77  & 3219.90 \\ 
20 & 3220.45  & 3228.68 \\ 
21 & 3282.26  & 3290.78 \\ 
22 & 3286.79  & 3304.68 \\ 
23 & 3322.10  & 3322.31 \\ 
24 & 3324.46  & 3335.16 \\
\bottomrule
\end{tabular*}
\end{table}

\section{Example Input Files}
\subsection{SHARC Input Files}
Example \SHARC \texttt{input} file for trajectory surface hopping dynamics of \textit{trans}-azomethane out to \qty{400}{\fs} with a nuclear time step of \qty{0.5}{fs}.
\begin{lstlisting}
printlevel 2

geomfile "geom"
veloc external
velocfile "veloc"

nstates 2
actstates 2
state 2 mch
coeff auto
rngseed 12118

ezero    -188.0849123900
tmax 400.000000
stepsize 0.500000
nsubsteps 200
integrator fvv

method tsh
surf mch
coupling ktdc
nogradcorrect
ekincorrect parallel_vel
reflect_frustrated none
decoherence_scheme edc
decoherence_param 0.1
hopping_procedure sharc
grad_all
select_directly
nospinorbit
output_format ascii
output_dat_steps 1
nophases_from_interface
ktdc_method gradient
\end{lstlisting}

\subsection{PySCF Template Files}
Example \texttt{PYSCF.template} file to perform L(2)-tPBE(6,4) with the 6-31G* basis set.
\begin{lstlisting}
basis 6-31g*
ncas 4
nelecas 6
roots 2

method l-pdft
pdft-functional tpbe
grids-level 4
conv-tol 1e-7
conv-tol-grad 1e-4

max-cycle-macro 10000
grad-max-cycle 1000

verbose 3
\end{lstlisting}

Example \texttt{PYSCF.template} file to perform SA(2)-CASSCF(6,4) with the 6-31G* basis set.
\begin{lstlisting}
basis 6-31g*
ncas 4
nelecas 6
roots 2

method casscf
conv-tol 1e-7
conv-tol-grad 1e-4

max-cycle-macro 10000
grad-max-cycle 1000

verbose 3
\end{lstlisting}

%\bibliography{supplement}